\documentclass[final,5p,twocolumn]{elsarticle}
\usepackage{lineno,hyperref}
\usepackage{dcolumn}
\modulolinenumbers[5]

\RequirePackage{graphicx}
\usepackage{dcolumn}
\usepackage{bm}
\usepackage{color}
\usepackage{amssymb}
\usepackage{epstopdf}
\usepackage{caption}
\usepackage{epsfig}
\usepackage{setspace}
\usepackage{amsbsy}
\usepackage{subfig}
\usepackage{physics}

\begin{document}

\begin{frontmatter}

\title{Classical Universe Arising from Quantum Cosmology}

\author{S. Jalalzadeh$^{1}$}
\ead{shahram.jalalzadeh@ufpe.br}

\author{M. Rashki$^{2}$}
\ead{mahdirashki@uoz.ac.ir}
\author{S. Abarghouei Nejad$^{1}$}
\ead{salmanabar@df.ufpe.br}
\address{$^{1}$Departamento de F\'{i}sica, Universidade Federal de Pernambuco,
Recife, PE, 52171-900, Brazil}
\address{$^{2}$Department of Physics,  University of Zabol, 98615-538 Zabol, Sistan and Balouchestan, Iran}



\begin{abstract}
In this paper, we study the classical limit {and unitary evolution} of quantum cosmology by applying the Weyl--Wigner--{Groenewold}--Moyal formalism of deformation quantization to quantum cosmology of a homogeneous and isotropic universe with positive spatial curvature and conformally coupled scalar field. The corresponding quantum cosmology {(similar to the Schr\"{o}dinger interpretation in canonical quantization scheme of quantum cosmology)} is described by the Moyal--Wheeler--DeWitt equation which has an exact solution in Moyal phase space, resulting in Wigner quasiprobability distribution function, peaking over the classical solutions. We show that for a large value of the quantum number $n$, the emerged classical universe is filled with radiation with quantum mechanical origin. Also, we introduce a canonical transformation on the scalar field sector of the model such that the conjugate momenta of the new canonical variable appear linearly in the transformed total Hamiltonian. Using this canonical transformation, we show that, it may lead to disentangle the time from the true dynamical variables.  We obtain the time-dependent Wigner function for a coherent as well as for squeezed states. We show that the peak of these Wigner functions follows the classical trajectory in the phase space.
\end{abstract}

\begin{keyword}
Quantum cosmology\sep Deformation quantization\sep Wigner function\sep Dirac and Friedmann observables\sep Coherent state \sep Time evolution
\end{keyword}

\end{frontmatter}


\section{Introduction}
{The canonical formalism of quantum cosmology} is based on the Wheeler--DeWitt (WDW) equation, which represents the wave function of the whole universe \cite{Hartle,Kiefer}. Hence, constructing wave functions {in the Schr\"{o}dinger interpretation} \cite{Isham1,Book}, obtained from the solutions of the WDW equation, is a suitable approach to investigate quantum cosmology. {Hereof, the square of the wave function of a solution of WDW equation is interpreted as the probability density for
``finding'' the cosmological model in a specific state
\begin{eqnarray}
\text{Prob}(R;\Psi)=\int_R\Psi^*(q)\Psi(q)\sqrt{-\mathcal G}d^Dq,
\end{eqnarray}
  where $R$ is a subset of minisuperspace under consideration and $\sqrt{-\mathcal G}d^Dq$ is the natural volume element in $D$-dimensional minisuperspace. The  Schr\"{o}dinger interpretation has been used in studies of quantum cosmology models by various authors, as an example see, \cite{Hartle:1983ai,HAWKING1984257,PhysRevD.39.2216}.
 In general, a conventional application of the  Schr\"{o}dinger interpretation is
to pose questions of the type ``What is the probability of finding this or that universe?'' or ``how quantum mechanical wave formalism has a classical description?''
rather than questions dealing with various evolution models of the same universe. } The generally accepted explanation is that quantum mechanics is the fundamental theory and the classical behavior is only a limit for large systems, i.e. a quantum system with enough large quantum number. Generalizing this scope, one may ask how the classical limit of our world can emerge from a quantum cosmology theory? Actually, it is not an easy question to answer. As a viewpoint one can regard the peaking of the wave function, providing a way to study the classical limit by preparing a comparison between the classical and quantum dynamics in the phase space. This implies that for a classical limit being existed, the quantum world should occur while peaking around a classical trajectory in the phase space \cite{Habib}.

But, while dealing with the wave function, which is obtained by WDW equation or path integral, a {conceptual} problem is arisen where it is necessary to know how to construct a proper wave packet, peaked around the original classical cosmological model \cite{Alvarenga}. In ordinary quantum mechanics, where one describes the dynamics of an ensemble of identical systems, the wave packet reduction in the Copenhagen interpretation leads to no practical problem. But in cosmology, the observer is an element of the universe itself, i.e. there is only one universe as a system. Therefore, the corresponding wave function of the state of the universe is not clear.

Another problem that arises here is called the measurement problem. During an observation in the standard quantum mechanics, the quantum system interacts with a classical domain where  the classical domain necessarily comes from how it solves the measurement problem \cite{Omnes}. In a conventional measurement, the wave function plus measuring device splits into non-overlapping branches, containing the measured system in an eigenstate of the measured observable, while the measuring device indicating the corresponding eigenvalue. Therefore, the wave function collapses into an eigenstate of the observable, and the other branches disappear. This is due to the short duration and strong coupling interaction between the measured system and the classical measuring device. In the Copenhagen interpretation, a real collapse of the wave function cannot be described by the unitary quantum evolution, and the fundamental measuring process should occur outside the quantum system, i.e. in a classical realm. But this is problematic, since in quantum cosmology, as a quantum theory of the whole universe, there is no place for a classical domain outside of that.
Hence, an improved scheme is needed to be applied to quantum cosmology. Some models such as de Broglie--Bohm interpretation of quantum cosmology \cite{Shojai}, quantum Hamilton-Jacobi cosmology \cite{Fathi}, and deformation quantization of cosmology \cite{Antonsen,Bayen}, are proposed to overcome the quantum cosmological difficulties about the measurement problem while maintaining the universality of quantum theory and the emergence of the classical universe.

As we know, one of the most important features which are related to the WDW equation and the corresponding Wigner function is the problem of time. The Hamiltonian of general relativity is a combination of first-class constraints and consequently, it vanishes on the reduced state space, which means that there is no notion of time evolution in the space with true degrees of freedom.
 The problem of the time was addressed for the first time by DeWitt. However, in a series of articles  \cite{DeWitt} he argued that this problem should not be considered as an obstacle in the sense that the theory itself would contain a suitable well-defined time parameter. In this scheme, ``intrinsic'' time parameter is identified with one of the variables of the $3D$ submanifold (usually the scale factors or the conjugate momenta corresponding to the scale factors), or with a scalar character of matter field(s) coupled to gravity, known as the ``extrinsic'' time. Different approaches, arising from these two methods to ``disentangle the time'' from the
``true dynamical variables'' have been investigated in detail in \cite{Time}.
If the selected time variable results in a close correspondence
between the expectation value of the dynamical variable and the classical prediction (prediction of general relativity) for long enough time, the selected time variable can be considered as acceptable.

{In this paper, we study the deformation quantization (which is also known as the phase space formalism of quantum mechanics) of quantum cosmology}. Deformation quantization is based on the Wigner quasi-distribution function and Weyl correspondence between quantum mechanical operators and ordinary phase space functions \cite{Wigner,weyl,Gerstenhaber}.
In this formalism, observables are not represented by operators and are defined as the functions of phase space variables. To get a quantum mechanical description, the algebra of phase space functions is changed via replacing common point-wise product
between observables with an associative, but noncommutative, pseudo-differential star-product \cite{Bayen,Groenewold,Moyal}. Thus, instead of changing the nature of classical phase space functions, deformation quantization only deforms the structure of the corresponding algebra.
Applying this approach to pass from classical to quantum cosmology has the advantage of making quantum cosmology calculations similar to the Hamiltonian formalism of classical cosmology, to stay away from doing arduous operator calculations \cite{Antonsen}.

In this article, by applying Arnowitt--Deser--Misner
(ADM) formalism  \cite{Arnowitt}, we construct the Wigner function of a conformally coupled scalar field in {the  na\"{i}ve and the internal time interpretations of quantum cosmology. In the first interpretation of the model, we try to obtain the classical-quantum correlation by finding the extremum of the quasi-distribution probability function.  One the other hand, in the second interpretation, by using the Hamiltonian of the scalar field part, we obtain the proper internal time ``coordinate'' to obtain a time-evolving picture of the Wigner function of the model. The comparison of these two pictures presents a complementary theme; i.e.  in the Schr\"{o}dinger   interpretation of the model (represented in Fig.(\ref{fig1})) one can measure the energy of the matter content of the universe (the scalar field here) with higher accurately while losing the time concept. One the other hand, in the internal time interpretation,  measuring the position time evolution of the universe accurately will disturb the corresponding conjugate momenta, representing the energy of the scalar field in the universe. }

This article is organized as follows. In section \ref{ch2}, we explain a classical model as a non-minimally coupling of a free scalar field with gravity with a positive curvature FLRW background. In section \ref{ch3}, we analyze the quantum cosmology of the model and find that the Wigner function of the model is made of two independent Laguerre functions. Furthermore, we check the compatibility of classical and quantum solutions. We show that the classical universe emerged is radiation dominated, and its entropy with quantum cosmology origin is estimated thereafter. Finally, we construct time-dependent Wigner functions corresponding to coherent and squeezed states. We show that the peak of these Wigner functions follows the corresponding classical solution in the phase space.

 \section{The Classical Model }\label{ch2}
 In the ADM formalism, the $4D$ spacetime $\mathcal M$ is split (or decomposed) into a family
of spacelike three-hypersurfaces $\Sigma_t$, and the spacetime curvature scalar is expressed  via  the curvature
$^{(3)}R$ of $\Sigma_t$, its induced metric $h_{ab}$ ($a,b=1,2,3$), its extrinsic curvature tensor $K_{ab}$, the lapse function $N$
and the shift vector $N^a$. The ADM action functional of a non-minimally coupled scalar field $\Phi$ in natural units, ($\hbar=c=1$), is given by \cite{Kief11}
\begin{eqnarray}\label{0}
& \hspace{-1cm} S=\displaystyle\int_{t_i}^{t_f}dt\displaystyle\int_{\Sigma_t}\Big[\frac{N\sqrt{h}}{2}\Big(\frac{M_\text{P}^2}{4\pi}-\zeta\Phi^2\Big)\Big(\,^{(3)}R+K_{ab}K^{ab} -K^2\Big)\nonumber \\
&  -2\sqrt{h}\zeta\Phi\dot\Phi K-2\sqrt{h}\zeta\Phi\Phi_a\Big(KN^a-\sqrt{h}h^{ab}N_{,b}\Big) \nonumber \\
& -\frac{N\sqrt{h}}{2}\left(\frac{-\dot\Phi^2}{N^2}+h^{ab}\Phi_a\Phi_b+V(\Phi)\right)\Big]d^3x,
\end{eqnarray}
 where $M_\text{P}=1/\sqrt{G}$ is the Planck mass and $\zeta$ is a dimensionless coupling constant which is valued as $\zeta=0$ for minimal coupling, and $\zeta=\frac{1}{6}$ for conformal ($V(\Phi)=0$) coupling \cite{Faraoni}. Here, we apply the non-minimal coupling value for $\zeta$  with $V(\Phi)=0$ to have a conformally invariant $\Phi$. In classical cosmology scenarios, a conformal (non-minimal) coupling is usually referred to  the coupling of a scalar field and the Ricci scalar \cite{Abreu}, whereas it could have different types of interactions in different cosmological themes \cite{Linde,Halliwell,Magana}. Indeed, the conformal coupling of the scalar field seems to be interesting for several reasons \cite{Barbosa,pedram,Barros,Malekolkalami}. For instance, it allows us to explore the exact solutions of simple models, and at the same time, it is rich enough to be considered as a significant modification of quantum cosmology \cite{Rostami,Moniz,shahram}. Moreover, having a model with a coupled scalar field and gravity in hand, one can provide a more precise explanation for the effects of the curvature on the very early Universe \cite{Blyth}.

Let us consider a classical model which is consisted  of a cosmological system, presented by action \eqref{0}, and  a FLRW minisuperspace model with a constant positive  curvature with the line element
\begin{equation}\label{line}
\begin{array}{cc}
ds^{2}=-N^{2}(t)dt^{2}+ a^{2}(t)\left[\frac{dr^2}{1-r^2}+r^2\Big(d\theta^{2}+\sin^{2}\theta d\phi^{2}\Big)\right],
\end{array}
\end{equation}
where, $a(t) $ is the scale factor and the lapse function is identified by $N(t)$. Assuming the scalar field to be homogeneous and isotropic, $ \Phi=\Phi(t)$, for a conformally coupled case we substitute the line element \eqref{line} into the action functional \eqref{0}, and rescaling lapse function as $\tilde N=\frac{N}{3\pi aM_\text{P}}$ and introducing new variables $x_{1}(t) := a(t), x_{2}(t) :=\frac{\sqrt{2}a(t)\Phi(t)}{
\sqrt{3}M_{\textsc{p}}}$ we get
\begin{equation}\label{action2a}
S= -\int dt\Big( \frac{M_{\textsc{p}}}{2 \tilde{N}}(\dot{x_1}^2-\dot{x_2}^2) + \frac{1}{2}M_{\textsc{p}} \omega^2 \tilde{N} (x_1^2 - x_2^2) \Big),
\end{equation}
with $\omega = 3\pi M_{\textsc{p}}$. To construct the Hamiltonian of the model, we consider the conjugate momenta of $\{x_1, x_2\}$ defined by
\begin{equation}
\Pi_1=-\frac{M_{\textsc{p}}}{\tilde{N}}\dot{x_1}, \qquad
\Pi_2=\frac{M_{\textsc{p}}}{\tilde{N}}\dot{x_2}.
\end{equation}
The corresponding Hamiltonian in terms of $2D$ minisuperspace $\{x_1, x_2\} $ is
\begin{equation}\label{hamilton}
 H=\tilde{N} \mathcal{H}:= \tilde{N} (\mathcal{H}_{1}-\mathcal{H}_{2}),
\end{equation}
where $\mathcal{H}_{1}$ and $\mathcal{H}_{2}$ are the superhamiltonians of the gravitational and scalar field parts respectively which are introduced by
\begin{eqnarray} \label{super}
\begin{array}{cc}
\mathcal{H}_{1}:=\frac{1}{2M_{\textsc{p}}}\Pi_{1}^2  +\frac{1}{2}M_{\textsc{p}}\omega^{2}x_{1}^2,   \\
\\
\mathcal{H}_{2}:=\frac{1}{2M_{\textsc{p}}}\Pi_{2}^2  +\frac{1}{2}M_{\textsc{p}}\omega^{2}x_{2}^2.
\end{array}
\end{eqnarray}
The lapse function $\tilde{N}$ acts as a Lagrange multiplier. The variation of the above action with respect to $\tilde{N}$ yields to the superhamiltonian constraint
\begin{equation} \label{super-H}
  \mathcal{H}\approx 0.
\end{equation}
In conformal time gauge fixing, $\tilde{N}=1$, the classical solutions of field equations are given by
\begin{equation}\label{10}
  x_{1}=a_{max} \sin(\omega t),\hspace{1cm}   x_{2}=\eta a_{max} \sin(\omega t+\delta),
\end{equation}
where $\eta=\pm 1$  is imposed via using the superHamiltonian constraint \eqref{super-H} and $a_{max}$ is the maximum value of the scale factor $ a(t)$. The above solution implies that the classical solution to be displayed as following trajectories
\begin{equation}
  x_{1}^{2}+x_{2}^{2}-2\eta \cos \delta x_{1} x_{2}=a_{max}^{2} \sin^{2} \delta.
\end{equation}
\section{Quantum cosmology and emerged classical Universe}\label{ch3}
In deformation quantization, observables are represented by phase space functions. Consequently the principal element of deformation quantization, that lays in the algebraic structure of the theory, is an associative and noncommutative pseudo-differential star-product \cite{Bayen,Kontsevich}.
The quasi-probability distribution Wigner function, which corresponds to the state of the system, is a prominent component of phase space quantization and allows us to calculate expectation values and probabilities \cite{wigner}. The Wigner function in a 2D-dimensional phase space $(x^i,\Pi_j),\,i,j=1,2,..,D$, is introduced by
 \begin{eqnarray}\label{WF}
 \begin{array}{cc}
W_n(x,\Pi)=C\displaystyle\int{dy^{D} e^{-i\Pi y} \psi_{n}^{\ast}(x-\frac{ y}{2})  \psi_{n}(x+\frac{ y}{2}) },\\
\\\displaystyle\int d^Dxd^D\Pi W_n(x,\Pi)=1,
\end{array}
 \end{eqnarray}
where $C$ is a constant and $\psi_n(x)$ denotes a general state. In addition, the concept of star-product was introduced by Gerstenhaber \cite{Gerstenhaber}. To apply it into quantum mechanics, we should consider the Moyal-Groenewold star product, $*_{\textsc{m}}$, of two observables, say  $f(x,\Pi)$ and $g(x,\Pi)$, on a Poisson manifold as
\begin{equation}\label{Moyal product}
\begin{array}{cc}
f(x,\Pi) *_{\textsc{m}} g(x,\Pi):= \\
\\f(x,\Pi) \exp{\frac{i}{2}(\overleftarrow{\partial_{x}}\overrightarrow{\partial_{\Pi}}-\overleftarrow{\partial_{\Pi}}\overrightarrow{\partial_{x}})}g(x,\Pi).
\end{array}
\end{equation}
To obtain the Moyal--Wheeler--DeWitt equation, we introduce the formal form of the Moyal star-product between superHamiltonian  $H(x,\Pi)$ in \eqref{hamilton} and the Wigner function $W(x,\Pi)$
\begin{equation} \label{WDM}
H(x,\Pi)\ast_{\textsc{m}} W(x,\Pi)=0,
\end{equation}
in which the ordinary product of the observables in phase space is replaced by the Moyal product. To be in the form of Bopp's shift formula \cite{Bopp}, the Moyal--Wheeler--DeWitt \eqref{WDM} becomes
\begin{equation} \label{WDM1}
H\left(x_{i}+\frac{i}{2}\vec\partial_{\Pi_{i}},{\Pi_{i}}-\frac{i}{2}\vec\partial_{x_i}\right)W(x_{i},\Pi_{i})=0,
\end{equation}
with $i=1,2$, stating two modes of the superHamiltonian \eqref{hamilton}. This is equivalent to
\begin{equation} \label{WDM2}
\begin{array}{cc}
\Big[\frac{({\Pi_{1}-\frac{i}{2}\partial_{x_1}})^2}{2M_{\textsc{p}}}  +\frac{M_{\textsc{p}} \omega^2}{2} (x_1 +\frac{i}{2}\partial_{\Pi_1})^2 -
\frac{({\Pi_{2}-\frac{i}{2}\partial_{x_2}})^2}{2M_{\textsc{p}}} \\
\\-\frac{M_{\textsc{p}} \omega^2}{2} (x_2 +\frac{i}{2}\partial_{\Pi_2})^ 2  \Big]W(x_{1},x_{2},\Pi_{1},\Pi_{2})=0.
\end{array}
\end{equation}
Since the Wigner function is a real valued function, one can separate the real and imaginary parts of \eqref{WDM2} and obtain two coupled partial differential equations with the real part identified as
 \begin{equation} \label{WDM4}
 \begin{array}{cc}
 \Big[\frac{1}{2 M_{\textsc{p}}}(\Pi_{1}^{2}-\Pi_{2}^{2})+\frac{M_{\textsc{p}}\omega^2}{2}(x_{1}^{2}-x_{2}^{2})- \frac{1}{8M_{\textsc{p}}} (\partial^{2}_{x_1}+\partial^{2}_{x_2} ) \\
 \\-\frac{M_{\textsc{p}} \omega^2}{8}(\partial^{2}_{\Pi_1}+\partial^{2}_{\Pi_2}) \Big]W(x_{1},x_{2},\Pi_{1},\Pi_{2})=0,
 \end{array}
 \end{equation}
and the imaginary part as
\begin{equation} \label{WDM5}
\begin{array}{cc}
 \Big[\frac{1}{2 M_{\textsc{p}}}\big( \Pi_{1}\partial_{x_1}-\Pi_{2}\partial_{x_2} \big)-\\
 \\\frac{M_{\textsc{p}}\omega^2}{2} \big( x_{1} \partial_{\Pi_1}- x_{2} \partial_{\Pi_2} \big)  \Big]W(x_{1},x_{2},\Pi_{1},\Pi_{2})=0.
 \end{array}
 \end{equation}
The imaginary part (\ref{WDM5}) enforces a special symmetry  that the Wigner function  depending only on the superHamiltonians \eqref{super}, $W=W(\mathcal H_1,\mathcal H_2)$ \footnote{Note that Eq.(\ref{WDM5}) implies that  $W=W(\mathcal H_1,\mathcal H_2, x_1\Pi_2+x_2\Pi_1)$. However, as we show in Eq.(\ref{WW11}), the relation between Wigner function and the wave function indicate that the Wigner function is a function only of $\mathcal H_1$ and $\mathcal H_2$.}. This leads to have the following relations,
\begin{eqnarray}
\begin{array}{cc}
 \partial_{x_i}^2f(\mathcal H_i)=M_\textsc{p}\omega^2\partial_{\mathcal H_i}f(\mathcal H_i)+M_\textsc{P}^2\omega^4x_i^2\partial^2_{\mathcal H_i}f(\mathcal H_i),\\
 \\
 \partial_{\Pi_i}^2f(\mathcal H_i)=\frac{1}{M_\textsc{p}}\partial_{\mathcal H_i}f(\mathcal H_i)+\frac{\Pi_i^2}{M_\textsc{P}^2}\partial^2_{\mathcal H_i}f(\mathcal H_i).
 \end{array}
\end{eqnarray}
Substituting them into \eqref{WDM4}, we obtain the following differential equation
 \begin{eqnarray} \label{WDM9}
 \begin{array}{cc}
 \Big[(\mathcal{H}_1-\frac{\omega^2}{4}\partial_{\mathcal{H}_1}-\frac{\omega^2}{4} \mathcal{H}_1 \partial_{\mathcal{H}_1}^2) -\\
 \\(\mathcal{H}_2-\frac{\omega^2}{4}\partial_{\mathcal{H}_2}-\frac{\omega^2}{4} \mathcal{H}_2 \partial_{\mathcal{H}_2}^2)\Big]W(\mathcal{H}_{1},\mathcal{H}_{2})=0,
 \end{array}
 \end{eqnarray}
which leads to the separable Wigner function
\begin{equation}\label{w.sep}
W(\mathcal{H}_{1},\mathcal{H}_{1})=W_1(\mathcal{H}_{1})W_2(\mathcal{H}_{2}),
\end{equation}
each of them satisfying
\begin{equation} \label{W1}
 \mathcal{H}_{i}\frac{d^{2} W_i(\mathcal{H}_{i})} {d{\mathcal{H}_{i}}^{2}}+ \frac{d W_i(\mathcal{H}_{i})}{d{\mathcal{H}_{i}}} - \big(\frac{4 }{\omega^2 }\mathcal{H}_{i}-\frac{4E}{\omega^2}\big)  W_i(\mathcal{H}_{i})=0,
\end{equation}
where  $E$ is a separating constant. The equations \eqref{W1} give the following solutions in terms of Laguerre polynomials, $ L_n $,
\begin{equation}\label{wu}
W_i(\mathcal{H}_{i})=\frac{(-1)^n}{\pi} e^{-\frac{2\mathcal{H}_i}{\omega}} L_n\Big(\frac{4\mathcal{H}_i}{\omega}\Big),\,\,\,\,i=1,2,
\end{equation}
with natural numbers $n$, and $E=\omega(n+\frac{1}{2})$.
Hence, the Wigner function \eqref{w.sep} is
\begin{eqnarray}\label{wuvt}
\begin{array}{cc}
 W_n(\mathcal{H}_{1},\mathcal{H}_{2})=\\\frac{1}{\pi^2}  e^{-\frac{2 }{\omega}(\mathcal{H}_{1}+\mathcal{H}_{2})} L_n\Big(\frac{4}{\omega}\mathcal{H}_{1} \Big)L_n\Big(\frac{4}{\omega}\mathcal{H}_{2}\Big).
 \end{array}
\end{eqnarray}
{Let us see how one can directly obtain the above Wigner function from its definition in Eq.(\ref{WF}). First, we need to obtain the wave function $\Psi(x_1,x_2)=\psi_1(x_1)\psi_2(x_2)$ of the model, with the total Hamiltonian operator $\hat{\mathcal H}=\hat{\mathcal H}_1-\hat{\mathcal H}_2=0$. The classical scale factor satisfies the inequality $x_1=a\geq 0$, which means that we deal with the spatial metric reconstruction problem in the WDW equation. In addition, to have a Hermitian and a self-adjoint extension for the superHamiltonian of the gravitational part, $\mathcal H_1$, we need to chose a suitable boundary conditions at $x_1=0$. This problem can be tackled in  several  different ways such as: 1) Imposing standard commutation relations for the pair of conjugate operators $(\hat x_1,\hat\Pi_1)$, $[\hat x_1,\hat\Pi_1]=i$, even
though we know this leads to a spectrum for the scale factor, $\hat x_1$, which is the entire real line \cite{Isham1}.   In fact, it follows that the conception of extended superspace of DeWitt \cite{DeWitt1}, which in the present special case consists of extending the range of $x_1$ values to $(-\infty,\infty)$. The Hilbert space of the gravitational part will be $L^2(\mathbb R,dx_1)$ with the operators defined in the
usual way as
\begin{eqnarray}\label{sp1}
\hat x_1:=x_1,\qquad \hat\Pi_1=-i\frac{\partial}{\partial x_1}.
\end{eqnarray}
 In this approach, the problem  is how to give some physical meaning to the negative
values of the scale factor. Usually, we choose the absolute value of the scale factor, $|a|$, as a physically meaningful quantity \cite{Bojo}.

2)   One can insist on using the Hilbert space $L^2(\mathbb R^+, da)$ of the wave functions that are defined on the half-line but keep the definitions (\ref{sp1}). In this case, the conjugate momenta $\hat\Pi_1$ is
no longer a self-adjoint operator but nevertheless, it is possible to find for the superHamiltonian $\hat{\mathcal H_1}$ a self-adjoint extension. A necessary and sufficient condition to have a self-adjoint $\hat{\mathcal H_1}$ is given by Robin boundary condition \cite{shahram} for the gravitational sector of the wave function
\begin{eqnarray}
\frac{1}{\psi(x_1)}\frac{\partial\psi(x_1)}{\partial x_1}\Big|_{x_1=0}=\beta,\qquad \beta\in\mathbb R,
\end{eqnarray}
where $\beta$ is an arbitrary real constant which has the dimension of  the inverse of length. The parameter $\beta$ thus characterizes a 1-parameter family of self-adjoint extensions of the $\hat{\mathcal H_1}$ on the half-line. The problem in this approach is to give a physical meaning to the new parameter $\beta$ in the theory: $\beta$ would  be  a  new fundamental constant of the theory \cite{Tipler}. 3)   Using a canonical transformation at the classical level to a new variable $\alpha$ defined
by $x_1:=\exp(\alpha)$ with the conjugate momenta $\Pi_\alpha=\exp(-\alpha)\Pi_1$. Now, $\alpha\in\mathbb R$ and it can therefore
be quantized as part of a conventional set of commutation relations, $[\hat\alpha,\hat\Pi_\alpha]=i$, using the Hilbert space $L^2(\mathbb R, d\alpha)$. Since in this article, we use the Moyal bracket deformation (or quantization) given by
\begin{eqnarray}\label{canonic1}
\{\{x_i,\Pi_j\}\}=i,
\end{eqnarray}
in which
\begin{eqnarray}\label{canonic2}
\begin{array}{cc}
\{\{f,g\}\}:=\frac{1}{i}(f*_\text{M}g-g*_\text{M}f)\\
\\
=
2f(x,\Pi) \sin{\frac{i}{2}(\overleftarrow{\partial_{x}}\overrightarrow{\partial_{\Pi}}-\overleftarrow{\partial_{\Pi}}\overrightarrow{\partial_{x}})}g(x,\Pi),
\end{array}
\end{eqnarray}
is the Moyal bracket, we should use the first   Schr\"odinger quantization method as mentioned above to obtain the wave function. The resulting wave function is the wave function of a oscillator-ghost-oscillator system in which \cite{Tucker}
\begin{eqnarray}\label{wave}
\begin{array}{cc}
\Psi_n(x_1,x_2)=\frac{\sqrt{M_\textsc{P}\omega}}{\sqrt{\pi}2^n n!}\\\times H_n(\sqrt{M_\textsc{P}\omega}x_1)H_n(\sqrt{M_\textsc{P}\omega}x_2)e^{-\frac{M_\textsc{P}\omega}{2}(x_1^2+x_2^2)},
\end{array}
\end{eqnarray}
where $H_n(x)$ is a Hermite polynomial of the $n$th order.
Inserting (\ref{wave}) into (\ref{WF}) and after a little calculations we find
\begin{eqnarray}\label{WW11}
\begin{array}{cc}
W_n(x_i,\Pi_j)=\\
\displaystyle e^{-\frac{2}{\omega}(\mathcal H_1+\mathcal H_2)}\sum_{k,k'=0}^\infty C_{n,k,k'}{\mathcal H_1}^{n-k}{\mathcal H_2}^{n-k'}.
\end{array}
\end{eqnarray}
This means that $W_n(x,\Pi)$ is a function only of $\mathcal H_1$ and $\mathcal H_2$ and can be written as $W_n(\mathcal H_1,\mathcal H_2)$, as we find in (\ref{WDM5}).
Using Talmi transformation \cite{Shlomo} one can obtain the analytic form of the Wigner function (\ref{wuvt}).}
To conclude this part, let us mention that deformation quantization maintains several advantages to deal with more involved obstacles in quantum cosmology \cite{Cordero}. For example, to treat cosmological models with nontrivial topology, boundary conditions or with curved minisuperspaces is more suitable. For these circumstances (as we saw in obtaining the wave function in the above where the configuration space was a half-line), the conventional canonical quantization could lead to the presence of non-Hermitian operators. This issue is avoided in the  Weyl--Wigner--Groenewold--Moyal formalism as a result of the use of classical functions instead of self-adjoint operators.

\begin{figure}
  \includegraphics[width=1\linewidth]{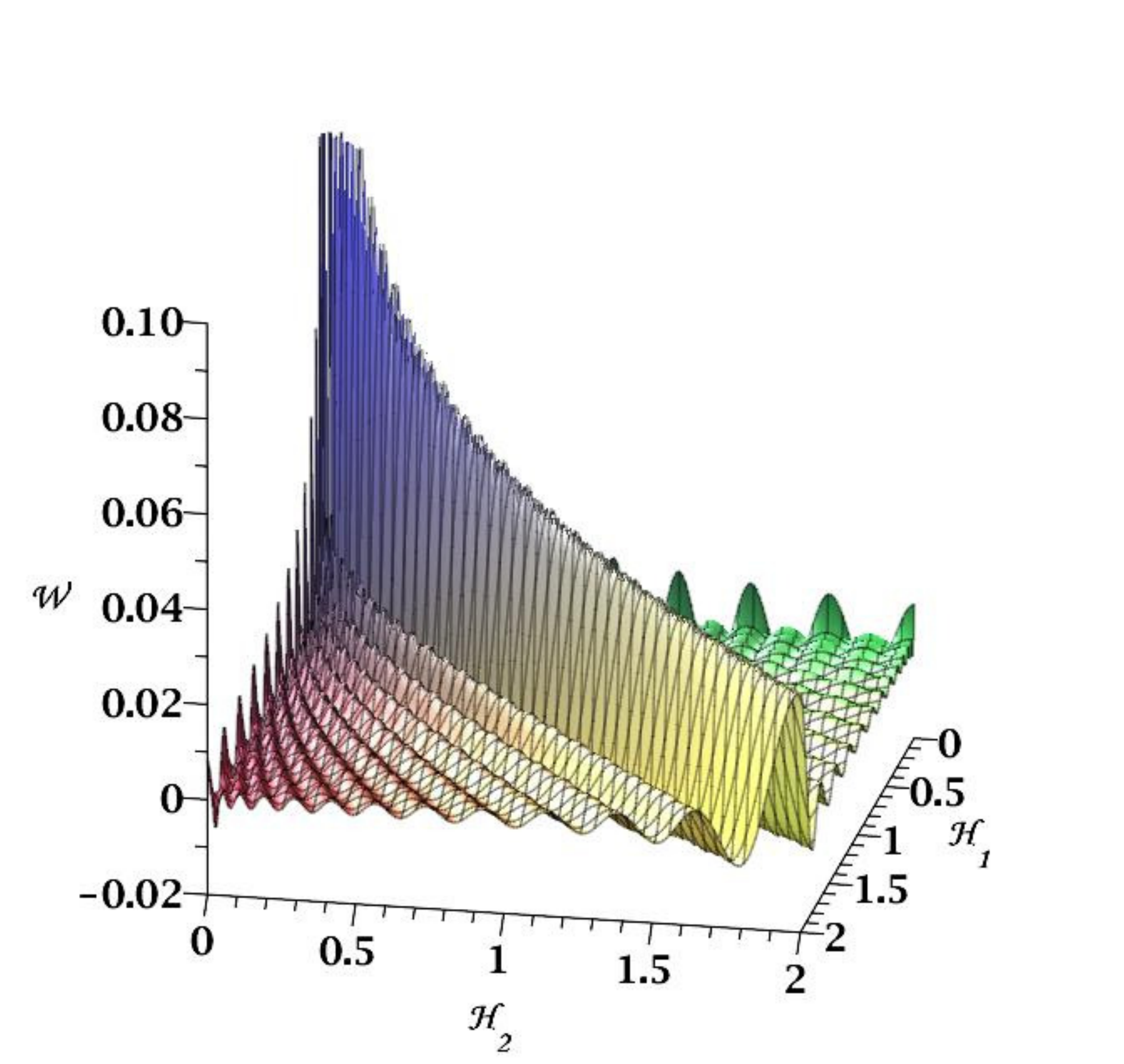}
  \caption{The Wigner function of a mixed state obtained in \eqref{new4} for $\omega =1$ and $N=220 $. The corresponding classical universe is $\mathcal H_1-\mathcal H_2=0$.}
  \label{fig1}
\end{figure}

Let us see how we can recover the corresponding classical cosmological model represented by superHamiltonian constraint \eqref{super-H}. In canonical quantization, resulted to the WDW equation, one needs usually to construct a coherent wave packet, with a suitable asymptotic behavior, peaking in the vicinity of the classical trajectory of the model in the minisuperspace \cite{Packet}. One may could extent this method to the Weyl--Wigner--Groenewold--Moyal formalism by using the Wigner function for the mixed states. To do this, let us consider a set of orthonormal pure states $\{|\psi_n\rangle, \langle\psi_m|\psi_n\rangle=\delta_{mn}\}$ with the corresponding Wigner functions $W_n$ obtained in (\ref{wuvt}). Then, from these Wigner functions of pure states we can construct a Wigner function for a mixed state as
\begin{eqnarray}\label{new1}
W=\sum_{n=0}^N c_nW_n,
\end{eqnarray}
where coefficients $c_i$ satisfy the following conditions \cite{Tatar}
\begin{eqnarray}\label{new2}
\sum_{n=0}^Nc_n=1,\,\,\,\,0\leq c_n\leq 1,\,\,\,\text{and}\,\,\,\sum_{n=0}^Nc_n^2<1.
\end{eqnarray}
For our subsequent analysis, by using Christoffel-Darboux formula \cite{Koe}
\begin{eqnarray}\label{new3}
\begin{array}{cc}
\displaystyle\sum_{n=0}^NL_n(x)L_n(y)=\\\begin{cases}
\frac{(N+1)}{x-y}\left(L_{N}(x)L_{N+1}(y)-L_{N+1}(x)L_N(y)\right),
\\\text{if}\,\,x\neq y,\\
\\
(N+1)\left(L_{N}(x)L^{(1)}_{N}(x)-L_{N+1}(x)L_{N-1}^{(1)}(x)\right),\\\text{if}\,\,x= y,
\end{cases}
\end{array}
\end{eqnarray}
where $L_n^{(\alpha)}(x)$ is generalized Laguerre polynomial, and substituting the Wigner function of the pure state (\ref{wuvt}) into (\ref{new1}), we can evaluate the sum over $n$ by choosing the coefficients $c_i$ as $c_1=c_2=...=c_N=1/N$
\begin{eqnarray}\label{new4}
\begin{array}{cc}
W(\mathcal H_1,\mathcal H_2)=\frac{(N+1)}{N\pi^2}\times\\
\begin{cases}
e^{-\frac{(x+y)}{2}}\frac{1}{x-y}\Big(L_{N}(x)L_{N+1}(y)-L_{N+1}(x)L_N(y)\Big),\\\text{if}\,\,x\neq y,\\
\\
e^{-x}\left(L_{N}(x)L^{(1)}_{N}(x)-L_{N+1}(x)L_{N-1}^{(1)}(x)\right),\\\text{if}\,\,x= y,
\end{cases}
\end{array}
\end{eqnarray}
where $x:=4\mathcal H_1/\omega$ and $y:=4\mathcal H_2/\omega^2$. Note that according to the correspondence principle, for large values of the $N$, we should expect a peak around the classical Hamiltonian of the system \cite{Berry}. Figure (\ref{fig1}) shows a plot of the Wigner function $W(\mathcal H_1,\mathcal H_2)$ obtained in (\ref{new4}) for $N=220$. It will be observed that among the small quantum fluctuations, there is a sharp extremum
in the vicinity of the classical loci $\mathcal H_1-\mathcal H_2=0$.
{
To understand the meaning of the above Wigner function, let us review the structure of the phase space of the model. According to Dirac \cite{Dirac}, an observable is a  quantity which has  vanishing Poisson bracket in the presence of the constraints. Regarding general relativity (GR), it is invariant under the group of diffeomorphism of hyperbolic spacetime manifold. Hamiltonian formalism of GR contains first-class constraints, namely, the Hamiltonian and momentum constraints. This leads to the conclusion that all GR Dirac observables should be time independent.
The unconstrained phase space of the model is $\mathbb R^4$ with global coordinates $(x_i,\Pi_i)$. However, all coordinates are not independent and they are connect by the Hamiltonian constraint (\ref{super-H}). Consequently, the physical phase space is a $3D$ subspace of $\mathbb R^4$ and all observables are confined to the physical phase space given by Hamiltonian constraint (\ref{super-H}).
If we form the quadratic combinations \cite{Rostami}
\begin{eqnarray}\label{Lie1}
\begin{cases}
J^{(i)}_+:=\frac{1}{2}(C_i^*)^2,\\
J^{(i)}_-:=\frac{1}{2}(C_i)^2,\\
J^{(i)}_0:=\frac{1}{2}C_iC_i^*=\frac{1}{2\omega}\mathcal H_i,\hspace{.5cm}i=1,2,
\end{cases}
\end{eqnarray}
  with
\begin{eqnarray}\label{Lie2}
\begin{array}{cc}
C_i:=\sqrt{\frac{M_\text{P}\omega}{2}}\left(x_i+\frac{i\Pi_i}{M_\text{P}\omega}\right),\\
C_i^*:=\sqrt{\frac{M_\text{P}\omega}{2}}\left(x_i-\frac{i\Pi_i}{M_\text{P}\omega}\right),
\end{array}
\end{eqnarray}
we obtain two Poisson $su(1,1)$ algebras
\begin{eqnarray}\label{Lie3}
\begin{cases}
\{J_0^{(i)},J^{(i)}_\pm\}=\mp iJ^{(i)}_\pm,\\
\\
\{J^{(i)}_+,J^{(i)}_-\}=2iJ^{(i)}_0.
\end{cases}
\end{eqnarray}
Furthermore, for
the gravitational and the scalar field parts, we define the central elements of the above algebras as
\begin{eqnarray}\label{Lie4}
\begin{array}{cc}
J_{(i)}^2:=(J_0^{(i)})^2-J_-^{(i)}J_+^{(i)}.
\end{array}
\end{eqnarray}
{The phase space functions $ \mathcal J \in  \{J_0^{(i)},J_+^{(i)},J_-^{(i)},J_{(i)}^2\}$, defined by (\ref{Lie1}) and (\ref{Lie4}),  are  all Dirac observables, since their Poisson brackets with the  superHamiltonian $\mathcal H = \mathcal H_1 - \mathcal H_2 $ vanish.
\begin{eqnarray}
\{\mathcal J,\mathcal H\}= 0.
\end{eqnarray}
}
 Thus, the extended phase space of the model is a $6D$ space spanned by $\Gamma=\{J^{(i)}_\pm,J^{(i)}_0,J_{(i)}^2\} $. On the other hand, superHamiltonian constraint (\ref{super-H}) indicate $J^{(1)}_0-J^{(2)}_0=0$ and $J^2_{(1)}=J^2_{(2)}=-3/16$, which show that only three of the set of Dirac observables are independent. Hence, the reduced phase space is $3D$ and one can choose three of  $\{J^{(i)}_0,J^{(i)}_\pm\}$ as a true coordinates of the phase space. At the quantum level, the quantum deformation of the Poisson algebra (\ref{Lie3}) is given by following two Lie--Moyal $su(1,1)$ algebras
\begin{eqnarray}\label{Lie5}
\begin{cases}
    \{\{J_0^{(i)},J^{(i)}_\pm\}\}=\pm J^{(i)}_\pm,\\
\\
\{\{J^{(i)}_+,J^{(i)}_-\}\}=-2J^{(i)}_0.
    \end{cases}
\end{eqnarray}
 This shows that at the quantum level (according to the uncertainty principle) the phase space is ``fuzzy'' and one cannot measure three-coordinates of a point on the phase space simultaneously.
The Wigner function (\ref{new4}) is expressed in terms of Dirac observables (or the coordinates) $J^{(i)}_0$
and it shows that the most probable universe is the original classical universe given by the Hamiltonian constraint (\ref{super-H}). The lake of information on the other two other coordinates, gives us the freedom to have a sharp pick around the classical solution which depends only $J^{(i)}_0$s. }

To see the classical-quantum correspondence in pure states, we use the large values of quantum number $n$ in pure Wigner functions (\ref{wuvt}). By using the asymptotic expansion of the Laguerre function, the Wigner function \eqref{wuvt} reduces to
\begin{equation}\label{classic.}
\begin{array}{cc}
W_n(\mathcal{H}_{1},\mathcal{H}_{2})\simeq\\ \frac{(-1)^n}{\pi^2 \omega^{\frac{1}{2}} \sqrt{n} \Big(\mathcal{H}_{1}\mathcal{H}_{2}\Big)^\frac{1}{4}} \cos{\big(4\sqrt{\frac{n \mathcal{H}_{1}}{\omega}}-\frac{\pi}{4}\Big)}\cos{\big(4\sqrt{\frac{n \mathcal{H}_{2}}{\omega}}-\frac{\pi}{4}\Big)}.\end{array}
\end{equation}
The locus of extremums of the above Wigner function
are given by the following simultaneous conditions
\begin{equation}\label{h class.}
\mathcal{H}_{1}=\frac{\pi^2 \omega}{16 n}\big( m_1 +\frac{1}{4}\big)^2 , \hspace{5mm}  \mathcal{H}_{2} = \frac{\pi^2 \omega}{16 n}\big( m_2 +\frac{1}{4}\big)^2,
\end{equation}
with $m_1,m_2 = 0,1,2,...\,$. These two superHamiltonians give us
\begin{equation}\label{h class.1}
\frac{{\Pi_{1}^2}}{2M_{\textsc{p}}}+\frac{1}{2}M_{\textsc{p}}\omega^{2}x_{1}^{2}-\frac{{\Pi_{2}^2}}{2M_{\textsc{p}}}-\frac{1}{2}M_{\textsc{pl}}\omega^{2}x_{2}^{2}-\mathcal{E}=0,
\end{equation}
where
\begin{equation}\label{e}
\mathcal{E} :=\frac{\omega \pi^2 (m_{1}-m_{2})}{32n}(1+2m_{1}+2m_{2}).
\end{equation}
{Hence, the extremums of the Wigner function of a pure state (\ref{classic.}) is given by (\ref{h class.1}) which  differs from the original Hamiltonian constraint (\ref{super-H}),  in that it contains  a constant $\mathcal E$, and it represents the semi-classical superHamiltonian for the most probable universes predicted by Wigner function (\ref{classic.}). In fact, the origin of $\mathcal E$ is uncertainty on measurement of superHamiltonians $\Delta\mathcal H_i$.}
Considering $ m_1 \simeq n \gg m_2 $, we get
\begin{eqnarray}\label{en}
\mathcal{E} = \frac{\pi^2 \omega n}{16}.
\end{eqnarray}
On the other hand, one can show that the quantum corrections are manifested in classical emerged universe as a perfect fluid of radiation type\footnote{The ADM action of a perfect fluid in the FLRW background is
\begin{eqnarray}
S=\frac{3V_kM_\textsc{P}^2}{8\pi}\int \Big(-\frac{a\dot a^2}{N}+kNa-\frac{8\pi}{3M_\textsc{P}}Na^3\rho\Big)dt,
\end{eqnarray}
where $\rho$ represents the energy density of the fluid and $V_k$ is the volume  of the spatial space with the constant curvature $k$. If we define the new lapse function (as we did in Eq.(\ref{action2a})) by $N=3V_kM_\textsc{P}a\tilde N/(4\pi)$, then the Hamiltonian of the model will be
\begin{eqnarray}
H_\textsc{ADM}=-\tilde N\Big(\frac{1}{2M_\textsc{P}}\Pi_a^2+\frac{k}{2}M_\textsc{P}\omega^2a^2-\frac{3M_\textsc{P}V_k^2}{4\pi}\rho a^4\Big),
\end{eqnarray}
where  $\Pi_a$ is the conjugate momenta to the scale factor and $\omega:=3V_kM_\textsc{P}/(4\pi)$. For radiation perfect fluid, $\rho=\rho_0a^{-4}$. Substituting the energy density of the radiation into the above ADM Hamiltonian gives us the superHamiltonian constraint equation
\begin{eqnarray}
\frac{1}{2M_\textsc{P}}\Pi_a^2+\frac{k}{2}M_\textsc{P}\omega^2a^2-\omega V_k\rho_0= 0,
\end{eqnarray}
where $\rho_0$ is the energy density of the fluid at the present epoch. Note that for $k=1$, the spatial volume is $V_k=2\pi^2$.}
as it is shown in \cite{rashki,rashki2}
\begin{equation}\label{e2}
\mathcal{E}=2\pi^2 \omega \rho x_{1}^4 = \frac{\omega}{8}\big(
\frac{1215}{\pi^4}\big)^{\frac{1}{3}}S_{\gamma}^{\frac{4}{3}},
\end{equation}
where $\rho$ and $S_{\gamma}$  are energy density and entropy of radiation, respectively. The equality of relations \eqref{en} and \eqref{e2} gives the estimation for the radiation entropy as $ S_\gamma \sim n^{\frac{3}{4}}$.
Using the current value of the entropy  of radiation, i.e. $ S_\gamma \sim 10^{88} $, we can estimate the approximate value of the quantum number $ n$ as $ n \sim 10^{117} $ which is in agreement with \cite{Moniz,rashki}.

Let us see how we can introduce a time-evolving Wigner function. To have a time-dependent Moyal--Wheeler--DeWitt equation, we introduce a canonical transformation on the scalar field sector of the Hamiltonian (\ref{hamilton}) such that in terms of
the new canonical variables, the total Hamiltonian contains linear momentum. Consider the following canonical transformation $(x_2,\Pi_2)\rightarrow(T,\Pi_T)$
\begin{eqnarray}\label{Time}
\begin{cases}
x_2:=\sqrt{\frac{2\Pi_T}{M_\text{P}\omega^2}}\sin(\omega T),\\
\Pi_2:=\sqrt{2M_\text{P}\Pi_T}\cos(\omega T).
\end{cases}
\end{eqnarray}
 Under this transformations, the Hamiltonian (\ref{hamilton}) takes the form
\begin{eqnarray}\label{T1}
H=\tilde N\left(\frac{1}{2M_{\textsc{p}}}\Pi_{1}^2  +\frac{1}{2}M_{\textsc{p}}\omega^{2}x_{1}^2-\Pi_T\right),
\end{eqnarray}
which shows that the coordinates of the reduced $3D$ phase space are $\Gamma=\{J^{(1)}_\pm,\Pi_T\}$.
The classical field equations then will be
\begin{eqnarray}\label{T2}
\begin{cases}
\dot x_1=\frac{\tilde N}{M_\text{P}}\Pi_1,\\
\dot\Pi_1=-M_\text{p}\omega^2\tilde Nx_1,
\end{cases}
\begin{cases}
\dot T=-\tilde N,\\
\dot\Pi_T=0.
\end{cases}
\end{eqnarray}
In conformal time gauge fixing, $\tilde{N}=1$, the solutions of the second set of the above dynamical equations are
\begin{eqnarray}\label{T3}
\Pi_T=const.,\qquad T=-t,
\end{eqnarray}
which show that $T$ plays the role of time parameter. The solutions of the first set in (\ref{T2}) are
\begin{eqnarray}\label{T4}
x_1=a_{max}\sin(\omega t),\qquad \Pi_1=M_\text{P}\omega a_{max}\cos(\omega t),
\end{eqnarray}
that are the classical solutions for scale factor obtained in (\ref{10}). {Note that the Poisson bracket of the time parameter $T$ does not vanish with superHamiltonian, $\{T,\mathcal H\}=\{T,\Pi_T\}=1$, which shows that $T$ is not a Dirac observable and so eligible to be considered as a time variable
to measure the passage of time \cite{Timepass}.}

{The canonical transformations (\ref{Time}) maps the phase space $(x_2,\Pi_2)$ onto the space $(T,\Pi_T)$,  meaning that the area element of the phase space $d\sigma'$ in the new coordinates is related to the old area element $d\sigma$ by $d\sigma=Jd\sigma'$, where $J$ is the Jacobian of the transformation. As a result, the Moyal--Groenewold star product defined by (\ref{Moyal product}) yields
\begin{eqnarray}\label{HH11}
\begin{array}{cc}
f(x_2,\Pi_2) *_{\textsc{m}} g(x_2,\Pi_2):= \\
\\
f(x_2,\Pi_2) \exp{\frac{i}{2}(\overleftarrow{\partial_{x_2}}\overrightarrow{\partial_{\Pi_2}}-\overleftarrow{\partial_{\Pi_2}}\overrightarrow{\partial_{x_2}})}g(x_2,\Pi_2)\\
\\=
f(T,\Pi_T) \exp{\frac{i}{2}J^{-1}(\overleftarrow{\partial_{T}}\overrightarrow{\partial_{\Pi_T}}-\overleftarrow{\partial_{\Pi_T}}\overrightarrow{\partial_{T}})}g(T,\Pi_T)\\
\\=
f(T,\Pi_T) *_{\textsc{m}} g(T,\Pi_T),
\end{array}
\end{eqnarray}
where in the last equality we used the fact that $J=1$. Hence, the Moyal--Groenewold star product is preserved under canonical transformation (\ref{Time}).}

Note that the new set of phase space coordinates $(T,\Pi_T)$ are related to the action-angle variables $(\varphi,p_\varphi)$ of harmonic oscillator by
\begin{eqnarray}\label{TTT}
T=\frac{\varphi}{\omega},\hspace{1cm}\Pi_T=\omega p_\varphi.
\end{eqnarray}
In terms of action-angle variables, the classical observables (\ref{Lie1}) for the scalar field part will be
\begin{eqnarray}\label{man1}
\begin{cases}
J^{(2)}_+=-\frac{1}{2}p_\varphi e^{2i\varphi},\\
J^{(2)}_-=-\frac{1}{2}p_\varphi e^{-2i\varphi},\\
J^{(2)}_0=\frac{1}{2}p_\varphi.
\end{cases}
\end{eqnarray}
{ In the new action-angle coordinates (\ref{Time}), the Moyal--Wheeler--DeWitt equation for the scalar field, $\mathcal H_2$, will be
\begin{eqnarray}\label{H2a}
\Pi_T\ast_{\text{M}}W_2=EW_2,
\end{eqnarray}
where $E$ is defined in Eq.(\ref{W1}). Using the Bopp's shift formula, the above equation turns to
\begin{eqnarray}\label{H2b}
\left(\Pi_T-\frac{i}{2}\frac{\partial}{\partial T}\right)W_2=EW_2.
\end{eqnarray}
Since $W_2$ is a real valued function, the above equation leads to $E=\Pi_T$ and $W_2=W_2(\Pi_T)$.
To obtain the explicit form of $W_2$, we solve the corresponding WDW equation. Thus, let us define the set of the action-angle (or angular) operators $\{e^{\pm i\hat\varphi},\hat p_\varphi\}$
\begin{eqnarray}\label{action1}
\begin{array}{cc}
e^{i\hat\varphi}. e^{-i\hat\varphi}=e^{-i\hat\varphi}. e^{i\hat\varphi}=1,\\
\hat p_\varphi=\hat p_\varphi^\dagger,\,\,\,(e^{i\hat\varphi})^\dagger=e^{-i\hat\varphi},\\
\qty[\hat p_\varphi,e^{i\hat\varphi}] =e^{i\hat\varphi},\\
\hat p_\varphi |n\rangle=n|n\rangle,\,\,\,n=0,\pm1,\pm2,...,\\
e^{\pm i\hat\varphi}|\varphi\rangle =e^{\pm i\varphi}|\varphi\rangle,\\
e^{\pm i\hat\varphi}|n\rangle=|n\pm1\rangle,
\end{array}
\end{eqnarray}
where $|n\rangle$ and $|\varphi\rangle$ are eigenvectors of $\hat p_\varphi$ and $e^{\pm i\hat\varphi}$, respectively.
The operator form of the canonical transformations (\ref{Time}) is given by \cite{Robert}
\begin{eqnarray}\label{action2}
\begin{cases}
\hat x_2:=-\frac{i}{\sqrt{2M_\text{P}\omega}}\Big(\sqrt{\hat p_\varphi}e^{i\hat\varphi}-e^{-i\hat\varphi}\sqrt{\hat p_\varphi}\Big),\\
\hat \Pi_2:=\sqrt{\frac{M_\text{P}\omega}{2}}\Big(\sqrt{\hat p_\varphi}e^{i\hat\varphi}+e^{-i\hat\varphi}\sqrt{\hat p_\varphi}\Big),
\end{cases}
\end{eqnarray}
where the square-root symbol is interpreted  as $\sqrt{\hat p_\varphi}|n\rangle=\sqrt{n}|n\rangle$. The Hermicity of $\hat x_2$ and $\hat\Pi_2$ implies $n$ have to be a non-negative integer, $n=0,1,2,...$ . In terms of the new variable, the superHamiltonian of scalar field part, $\mathcal H_2$ will be
\begin{eqnarray}\label{action3}
\hat{\mathcal H}_2=\omega(\hat p_\varphi+\frac{1}{2}).
\end{eqnarray}
Hence, the wave function of $\mathcal H_2$ is given by \cite{Oliveira}
\begin{eqnarray}\label{H2c}
&& \psi_2(\varphi)=\langle\varphi|n\rangle=\frac{1}{\sqrt{2\pi}}e^{i(n+\frac{1}{2})\varphi} \nonumber \\
&& \hspace{1cm} =\frac{1}{\sqrt{2\pi}}e^{i\omega(n+\frac{1}{2})T},\,\,\,n=0,1,2,...\,,
\end{eqnarray}
where we used $\varphi=\omega T$ in the last equality.
  Substituting  (\ref{H2c}) into Eq.(\ref{WF}) and doing a simple calculation, we get
\begin{eqnarray}\label{H2d}
\begin{array}{cc}
W_2(\Pi_T)=\\\delta\left(\omega p_\varphi-\omega(n+\frac{1}{2})\right)=
\delta\left(\Pi_T-\omega(n+\frac{1}{2})\right),
\end{array}
\end{eqnarray}
where we used $\Pi_T=\omega p_\varphi$ in the last equality.
This is essentially a statement of Bohr--Sommerfeld quantization: for state $|n\rangle$, the phase space distribution in
action-angle variables is that the classical action $\Pi_T$ is the quantum
half-integer $n$, and the angle variable $\omega T$ is random (i.e., the distribution is independent of $T$).}
By  introducing the  new form of the Hamiltonian in (\ref{T1}), the Moyal--Wheeler--DeWitt Eq. (\ref{WDM1}) will turn to the following Moyal--Wheeler--DeWitt--Schr\" odinger equation
\begin{equation} \label{T5}
\left(\frac{1}{2M_\text{P}}\Pi_1^2+\frac{1}{2}M_\text{P}\omega^2x_1^2-\Pi_T \right)*_\text{M}W=0.
\end{equation}
One may rewrite the above equation in terms of Bopp's shif formula
\begin{eqnarray}\label{T5a}
\begin{array}{cc}
\Big(\frac{1}{2M_\text{P}}(\Pi_1+\frac{i}{2}\partial_{x_1})^2+\\\frac{1}{2}M_\text{P}\omega^2(x_1-\frac{i}{2}\partial_{\Pi_1})^2-\Pi_T+\frac{i}{2}\partial_T \Big)W=0.
\end{array}
\end{eqnarray}
 {In differential equation (\ref{T5a}), the Wigner function is only function of Friedmann observables  ($x_1,\Pi_1$). All other Friedmann observables (for example,  redshift, density parameter and deceleration parameter) are functions  of these coordinates of $2D$ reduced phase space \cite{Timepass}.}
 Using  $\Pi_T=E$ (as we find in (\ref{H2b})) and separating the real and imaginary parts (remember $W$ is a real function) of the above differential equation gives us the following two differential equations
 \begin{eqnarray}\label{T5b1}
 \left(\mathcal H_1-\frac{1}{8M_\text{P}}\partial^2_{x_1}-\frac{M_\text{P}\omega^2}{8}\partial_{\Pi_1}^2 \right)W=EW,
 \end{eqnarray}
 and
 \begin{eqnarray}\label{T5b2}
 \partial_TW=\left(M_\text{P}\omega^2x_1\partial_{\Pi_1}-\frac{1}{M_\text{P}}\Pi_1\partial_{x_1} \right)W.
 \end{eqnarray}
 As we find in (\ref{WDM4}), Eq.(\ref{T5b1}) presents differential equation for Wigner function of a simple harmonic oscillator with mass $M_\text{P}$. On the other hand, Eq.(\ref{T5b2}) shows the time dependency of the Wigner function.
The solution of (\ref{T5b2}) is \cite{Zac}
\begin{eqnarray}\label{T6}
\begin{array}{cc}
W(x_1,\Pi_1,t)=\\
\\\exp{\left(M_\text{P}\omega^2x_1\partial_{\Pi_1}-\frac{\Pi_1}{M_\text{P}}\partial_{x_1}\right)T}W(x_1,\Pi_1,0),
\end{array}
\end{eqnarray}
where $W(x_1,\Pi_1,0)$ is the solution of (\ref{T5b1}).  The above solution shows a rotation around the origin in phase space. In matrix form, this rotation can be represented as
\begin{eqnarray}\label{T6a}
\begin{pmatrix}
\sqrt{M_\text{P}}\omega x_1'\\ \frac{1}{\sqrt{M_\text{P}}}\Pi_1'
\end{pmatrix}
=
\begin{pmatrix}
\cos(\omega t)&-\sin(\omega t)\\ \sin(\omega t)&\cos(\omega t)
\end{pmatrix}
\begin{pmatrix}
\sqrt{M_\text{P}}\omega x_1\\ \frac{1}{\sqrt{M_\text{P}}}\Pi_1
\end{pmatrix}.
\end{eqnarray}
The time evolution of the Wigner function lies in the phase space coordinates, so that it maintains its shape, but moves around in elliptical orbits in phase space. This means that
\begin{eqnarray}\label{T7}
\begin{array}{cc}
W(x_1,\Pi_1,t)=W\Big(x_1\cos(\omega t)\\-\frac{\Pi_1}{M_\text{P}\omega}\sin(\omega t),\cos(\omega t)\Pi_1+M_\text{P}\omega x_1\sin(\omega t),0\Big),
\end{array}
\end{eqnarray}
 rotates uniformly on the phase space and it is essentially classically, even though it provides a complete quantum mechanical description \cite{Zac}.
Let us take the Wigner function at time $t=0$ to be the lowest state of the harmonic oscillator (obtained in (\ref{wu})) shifted by $b=M_\text{P}\omega a_{max}$ in the $\Pi_1$ direction
\begin{eqnarray}\label{zero}
W(x_1,\Pi_1,0)=\frac{1}{\pi}\exp\left(-L^2(\Pi_1-b)^2-\frac{x_1^2}{L^2}\right),
\end{eqnarray}
where $L:=1/\sqrt{M_\text{P}\omega}=L_\text{P}/\sqrt{3\pi}$ in which $L_\text{P}=1/M_\text{P}$ is the Planck length. This is in fact, the Wigner function of a coherent state.
The Wigner function at other times becomes
\begin{eqnarray}\label{T8}
\begin{array}{cc}
W(x_1,\Pi_1,t)=\frac{1}{\pi}\exp\Big(-L^2(\Pi_1-b\cos(\omega t))^2\\-\frac{1}{L^2}(x_1-bL^2\sin(\omega t))^2\Big).
\end{array}
\end{eqnarray}
\begin{figure}
  \includegraphics[width=0.8\linewidth]{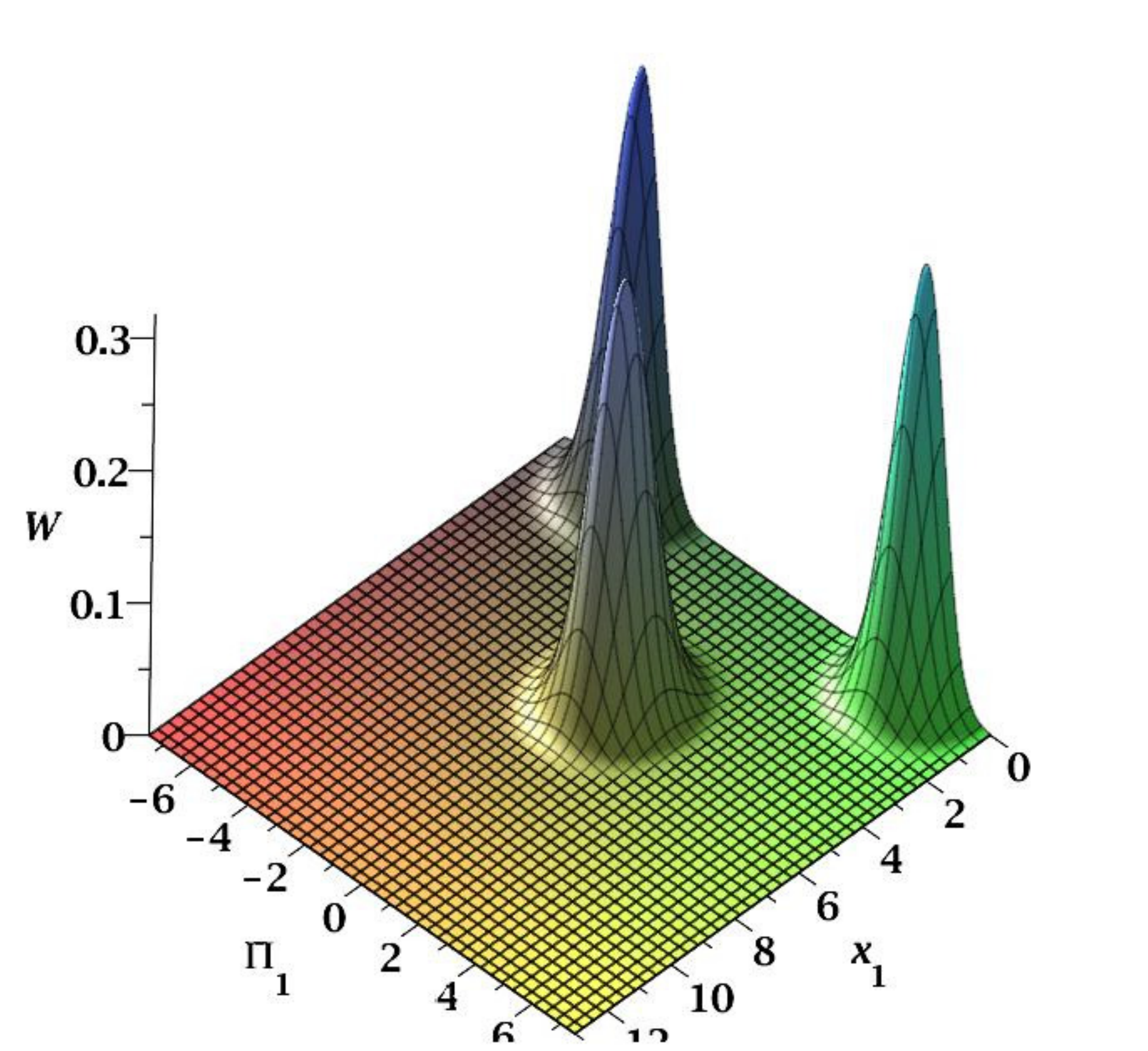}
  \caption{ Plot of the Wigner functions of a coherent state. The state start at the Big Bang, $t=0, x_1=0$, and moves in a clockwise fashion about the origin and finally it finish at the Big Crunch $t=\frac{\pi}{\omega},x_1=0$. It is shown at times $t=0$, $t=\frac{\pi}{2\omega}$, and $t=\frac{\pi}{\omega}$. In generating this plot the following values were used: $L_\text{P}=1$, $b=5$.}
  \label{fig2}
\end{figure}
\begin{figure}
  \includegraphics[width=0.8\linewidth]{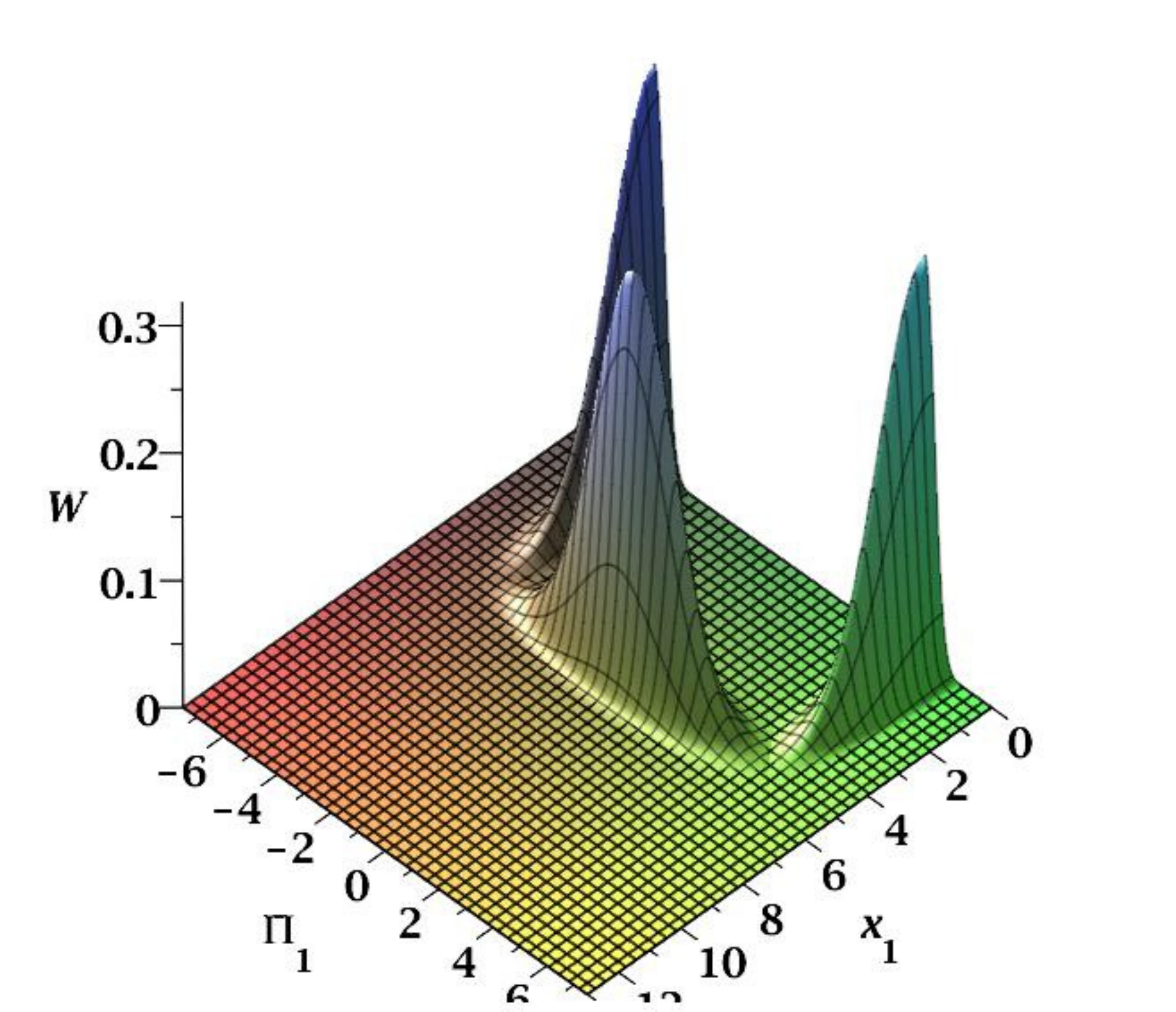}
  \caption{Plot of the Wigner functions of a squeezed state. The state start at the Big Bang, $t=0, x_1=0$, and moves in a clockwise fashion about the origin and finally it finish at the Big Crunch $t=\frac{\pi}{\omega},x_1=0$. It is shown at times $t=0$, $t=\frac{\pi}{2\omega}$, and $t=\frac{\pi}{\omega}$. In generating this plot the following values were used: $L=2$, $b=5$.}
  \label{fig3}
\end{figure}
As we see in Fig. \ref{fig2}, the time evolution
of the state is a motion in an ellipse in the $(x_1,\Pi_1)$ plane centered about the origin: the Wigner function  starts   at the Big Bang, $t=0$, and moves in a clockwise fashion  about  the origin of phase space, reaches to the point $(a_{max},\Pi_1=0)$ at $t=\frac{\pi}{2\omega}$ and finally it   finishes  at the Big Crunch, $t=\frac{\pi}{\omega},x_1=0$.
As the Wigner function moves
around its path in phase space, its projection on the scale factor  axis, $x_1$,  moves from the Big Bang towards the Big Crunch with unchanging profile. Note that in Wigner function (\ref{T7}), $L$  can be arbitrarily chosen, and generally it is not needed to be constrained by the relation $L=1/\sqrt{\omega M_\text{P}}$. In the later case, the Wigner function is
\begin{eqnarray}\label{last}
\begin{array}{cc}
W=\frac{1}{\pi}\exp\Big(-\frac{1}{L^2}\left(x_1\cos(\omega t)-\frac{\Pi_1}{\omega}\sin(\omega t)\right)^2-\\
L^2\left(\Pi_1\cos(\omega t)+M_\text{P}\omega x_1\sin(\omega t)-b\right)^2\Big),
\end{array}
\end{eqnarray}
and it
is the Wigner function of a squeezed state which have a different ratio of spread
in the scale actor and its momenta directions from the coherent state.  Fig.(\ref{fig3}) shows the Wigner function of a squeezed state. We find that  its projection on
the $x_1$ axis differs from that of the coherent state. Although
it will move from the Big Bang to the Big crunch, its width in $x_1$ will vary during the motion.

{To conclude, let see what is the relation of the Fig.(\ref{fig1}) and subsequent two other figures (\ref{fig2}) and (\ref{fig3}). In fact, they show the behavior of  Dirac and Friedmann Wigner function as a function of corresponding observables, respectively. Unlike the Dirac observables, Friedmann observables (like the age of the universe, Hubble parameter, density parameters, redshift, and so on) are time-dependent observables.
In Fig.(\ref{fig1}), for each point on the extremum of the Wigner function, the coordinates are given by $\mathcal H_1$ and $\mathcal H_2=\Pi_T$ and we have a minimum uncertainty for those coordinates.  The time coordinate $T$ and the corresponding momenta $\Pi_T$ satisfy Moyal bracket relation $\{\{T,\Pi_T\}\}=i$, consequently, minimizing the uncertainty on $\Pi_T=\mathcal H_2$ extremize the uncertainty on the time parameter $T$.  In other words, one can say that our universe is located on the peak of Fig.(\ref{fig1}) and the figure does not give us any information about the time. On the other hand for coherent and squeezed states (which realize the Friedmann observables), we have minimum uncertainty for the time, yielding a maximum uncertain value for $\Pi_T$ as a Dirac observable.  Thus, these figures, in one hand Fig(\ref{fig1}) and on the other hand figures \ref{fig2} and \ref{fig3}, are related to each other by two kind of observables, $\Pi_T$ and $T$, and the canonical commutation relation between them, $\{\{T,\Pi_T\}\}=1$. $\Pi_T$ and $T$ are complementary observables and as a result, Dirac and Friedmann observables of the model   are of the complementary ones}.

\section{Conclusion}\label{ch4}
{The problem of time in quantum cosmology has been studied from different points of view (e.g. see \cite{Isham1} and references therein). Among them, one can mention two main approaches, namely the internal time interpretation (in which a canonical transformation is used to obtain a Hamiltonian, linear in a variable such that its canonical conjugate can be used as a time variable), and the Schr\"odinger interpretation (belonging to the time-less category of quantum cosmology models). There is a vast literature, studying those aforementioned interpretations using the Wheeler--DeWitt equation. However in this paper, we used the Wigner function to compare two aforementioned approaches, and as a result, we showed that there is a complementary relation between them. In detail, in the Schr\"odinger interpretation of the model one can measure the energy of the matter content of the universe (the scalar field here) with higher accurately while losing the time concept, whereas in the internal time interpretation,  having the time evolution of the universe will disturb the corresponding conjugate momenta, representing the energy of the scalar field.}

By studying quantum cosmology of a conformally coupled scalar field in a positive curvature background of an FLRW type universe. We have solved the MWDW equation exactly and obtained four  types of Wigner functions,  regarding mixed, pure, coherent, and squeezed states. The quasi-probability distribution Wigner function is related to classical solutions without recourse to WKB
approximation techniques. The Wigner function of the mixed states for Dirac observables is plotted in Fig.(\ref{fig1}), showing that there exists a peak over the classical trajectory, indicating a good coincidence between classical and the most probable quantum states.   We showed that for large values of quantum number $n$ of the Wigner function of the pure states, the classical universe arose from the quantized model includes a radiation perfect fluid.
 Afterward, we introduced an extrinsic time parameter in which the resulting Wigner function is time-dependent. We introduced coherent state Wigner function in which as the Wigner function moves
around its path in phase space, its projection on the scale factor axis
moves from Big Bang to the Big Crunch (see Fig.\ref{fig2}) with an unchanging profile. Besides, we establish a squeezed Wigner function which represents an exact classical-quantum correlation in any time-evolution of the universe, which is plotted in Fig.(\ref{fig3}).



\begin{thebibliography}{99}
{\bibitem{Hartle} J.B. Hartle, T. Piran, S. Weinberg (Eds.), \textit{Quantum Cosmology and Baby Universes}, World Scientific, Singapore    (1991); L.J. Garay, J.J. Halliwell, G.A.M. Marugn, Phys. Rev. D \textbf{43}, 2572 (1991);
  S.P. Kim, Phys. Rev. D \textbf{46}, 3403 (1992);
  C. Barcel, M. Visser, Phys. Lett. B \textbf{466}, 127 (1999), arXiv:gr-qc/0003025;
  C. Kiefer, Nucl. Phys. B \textbf{341}, 273 (1990);
  G.D. Barbosa, Phys. Rev. D \textbf{71}, 063511 (2005),  arXiv:hep-th/0408071;}
  S. Capozziello, G. Lambiase, Gen. Relativ. Gravit.  \textbf{32}, 673 (2000), arXiv:gr-qc/9912083.

\bibitem{Kiefer} C. Kiefer; \textit{Quantum Gravity}, Oxford University Press, Oxford (2012);
  C. Kiefer, Ann. Phys. \textbf{15}, 129 (2005), arXiv:gr-qc/0508120;
  M. Bojowald, C. Kiefer, P.V. Moniz, arXiv:gr-qc/1005.2471;
  P.V. Moniz, \textit{Quantum Cosmology, The Supersymmetric Perspective}, Lecture Notes in Physics \textbf{803}, Springer-Verlag, Berlin (2010).


\bibitem{Isham1}  C.J. Isham, In Integrable Systems, Quantum Groups and Quantum Field Theories,
L.A. Ibort and M. A. Rodriguez (eds.) (Kluwer, London, 1993); C.J. Isham, J. Math. Phys. 23, 2157 (1994); C.J. lsham and N. Linden, J. Math. Phys. 35, 5452 (1994).

\bibitem{Book} P.V. Moniz and S. Jalalzadeh, \textit{Challenging Routes in Quantum Cosmology} World Scientific (2021).

\bibitem{Hartle:1983ai} J. B. Hartle and S. W. Hawking, Phys. Rev. D 28, 2960 (1983).

\bibitem{HAWKING1984257} S.W. Hawking, Nucl. Phys. B 239, 257 (1984).

\bibitem{PhysRevD.39.2216} M. Castagnino, Phys. Rev. D 39, 2216 (1989).


\bibitem{Habib} S. Habib, Phys. Rev. D 42, 2566 (1990).



\bibitem{Alvarenga} F.G. Alvarenga, J.C. Fabris, N.A. Lemos, G.A. Monerat, Gen. Rel. Grav. \textbf{34}, 651 (2002), arXiv:gr-qc/0106051;
N.A. Lemos, J. Math. Phys. \textbf{37}, 1449 (1996), arXiv:gr-qc/9511082; G.A. Monerat, G. Oliveira-Neto, E.V. Correa Silva, L.G. Ferreira
Filho, P. Romildo Jr., J.C. Fabris, R. Fracalossi, F.G. Alvarenga, S.V.B. Gonalves, Phys. Rev. D \textbf{76}, 024017
(2007), arXiv:gr-qc/0704.2585v1; F.G. Alvarenga, A.B. Batista, J.C. Fabris, Int. J. Mod. Phys. D \textbf{14}, 291 (2005), arXiv:gr-qc/0404034;
P. Pedram, S. Jalalzadeh, Phys. Lett. B \textbf{659}, 6 (2008),  arXiv:gr-qc/0711.1996; P. Pedram, S. Jalalzadeh, S.S.
Gousheh, Phys. Lett. B \textbf{655}, 91 (2007), arXiv:gr-qc/0708.4143; P. Pedram, S. Jalalzadeh, Phys. Rev. D \textbf{77}, 123529
(2008), arXiv:gr-qc/0805.4099.

\bibitem{Omnes} R. Omn\'es, The Interpretation of Quantum Mechanics, Princeton University Press, Princeton, (1994).


\bibitem{Shojai} F. Shojai, A. Shirinifard, Int. J. Mod. Phys. D \textbf{14}, 1333 (2005), arXiv:gr-qc/0504138; P. Pedram, S. Jalalzadeh, Phys. Lett. B \textbf{660}, 1 (2008), arXiv:gr-qc/0712.2593; P. Peter, N. Pinto-Neto, Phys. Rev. D \textbf{78}, 063506 (2008), arXiv:gr-qc/0809.2022.

\bibitem{Fathi} M. Fathi, S. Jalalzadeh, P.V. Moniz, Eur. Phys. J. C \textbf{76}, 527 (2016), arXiv:1609.04488.

\bibitem{Antonsen} F. Antonsen, Phys. Rev. D \textbf{56}, 920 (1997), arXiv:hep-th/9701182; H. Quevedo, J.G. Tafoya, Gen. Relativ. Gravit. \textbf{37}, 2083 (2005); H. Garc\'ia-Compe\'an, F.J. Turrubiates, Int. J. Mod. Phys. A \textbf{26}, 5241 (2011), arXiv:1109.1036; {
M. Rashki, S. Jalalzadeh, Phys. Rev. D \textbf{91}, 023501 (2015), arXiv:1412.3950; M. Rashki, S. Jalalzadeh, Gen. Rel. Grav. \textbf{49}, 14 (2017), arXiv:1612.03047.}

{\bibitem{Bayen} F. Bayen , M. Flato, C. Fronsdal, A.  Lichnerowicz,  D. Sternheimer, Annals Phys. \textbf{111}, 61 (1978).}
\bibitem{DeWitt} B.S. DeWitt, Phys. Rev. \textbf{160}, 113 (1967); Phys. Rev. \textbf{162}, 1195 (1967); Phys. Rev. \textbf{162}, 1239 (1967).
\bibitem{Time} J. Wess, B. Zuroino, Nucl. Phys. B\textbf{70} , 39 (1974); D. Volkov, V.P. Akuiov, Phys. Lett. \textbf{46B}, 109 (1973); J.D. Brown, K.V. Kucha\v r, Phys. Rev. D \textbf{51}, 5600 (1995);
M. Castagnino, Phys. Rev. D \textbf{39}, 2216 (1988);
J. Greensite, Nucl. Phys. B \textbf{342}, 409 (1990);
C.J. Isham, J. Butterfield, On the emergence of time in quantum gravity, arXiv:gr-qc/9901024;
R.M. Wald, Phys. Rev. D \textbf{48}, R2377 (1993);
C. Kiefer, Does time exist in quantum gravity?, arXiv:0909.3767;
E. Anderson, Problem of time in quantum gravity, arXiv:1206.2403;
A. Davidson, B. Yellin, Restoring time dependence into quantum cosmology, arXiv:1206.0830;  C. Teitelboim, ``Hamiltonian formulation of general relativity'', in Proceedings, Quantum cosmology and baby universes (Jerusalem 1989), 1-63.

{\bibitem{Wigner} E. Wigner, Phys. Rev. \textbf{40}, 749, (1932).}

{\bibitem{weyl} H. Weyl, Z. Phys. \textbf{46}, 1 (1927);
  H.  Weyl, \textit{Group Theory and Quantum Mechanics}, Dover, New York (1931).}

{\bibitem{Gerstenhaber} M. Gerstenhaber, Ann. Math. \textbf{79}, 59 (1964).}
{\bibitem{Groenewold} H. Groenewold, Physica \textbf{12}, 405 (1946);  G. Baker, Phys. Rev. \textbf{109},  2198 (1958);}

{\bibitem{Moyal} J. Moyal, Proc. Cambridge Philos. Soc. \textbf{45}, 99 (1949).}

{\bibitem{Arnowitt} R. Arnowitt, S. Deser, C.W. Misner, \textit{Gravitation: An Introduction to Current Research}, John Wiley, New York (1962).}

\bibitem{Kief11} C. Kiefer, Phys. Lett. B \textbf{225} (1989) 227.
\bibitem{Faraoni} S. Sonego, V. Faraoni, Classical. Quantum. Gravity \textbf{10}, 1185 (1993);
  V. Faraoni, Phys. Rev. D \textbf{53}, 6813 (1996), arXiv:astro-ph/9602111;
  V. Faraoni, Gen. Rel. Grav. \textbf{29}, 251 (1997) [arXiv:gr-qc/9608067];
  A.A. Grib, W.A. Rodrigues, Grav. Cosmol. \textbf{1}, 273 (1995).
{\bibitem{Abreu} P. Abreu, P. Crawford, J.P. Mimoso, Class. Quantum Grav. \textbf{11},  1919 (1994), arXiv:gr-qc/9401024;
    N. Cruz,  C. Martine, Class. Quantum Grav. \textbf{17},  2867 (2000), arXiv:gr-qc/9401024;
    S. Fay, Class. Quantum Grav. \textbf{21},  1609 (2004), arXiv:gr-qc/0402104;}

{\bibitem{Linde} A.D. Linde, \textit{Elementary Particle Physics and Inflation Cosmology}, Nauka, Moscow, (1990).}


{\bibitem{Halliwell} J.J. Halliwell, Phys. Lett. B \textbf{185}, 341 (1987).}

{ \bibitem{Magana} J. Maga\~na, T. Matos, J. Phys.: Conf. Ser. \textbf{378}, 012012 (2012), arXiv:1201.6107; M.A. Rodr\^iguez-Meza, Adv. Astron. \textbf{2012}, 509682 (2012), arXiv:1112.5201.}

{ \bibitem{Barbosa} G.D. Barbosa, Phys. Rev. D \textbf{71}, 063511 (2005), arXiv:hep-th/0408071.}

\bibitem{pedram} P. Pedram, Phys. Lett. B \textbf{671}, 1 (2009),  arXiv:gr-qc/0811.3668.

\bibitem{Barros} J.A. De Barros, N. Pinto–Neto, I.L. Shapiro, Class. Quant. Grav. \textbf{16}, 1773 (1999), arXiv:gr-qc/9809079.

\bibitem{Malekolkalami} B. Malekolkalami, M. Farhoudi, Phys. Lett. B \textbf{678}, 174 (2009), arXiv:0911.2548.

\bibitem{Rostami} T.  Rostami,   S. Jalalzadeh, P.V.   Moniz, Phys. Rev.D \textbf{92},  023526  (2015), arXiv:1507.04212.

\bibitem{Moniz} S. Jalalzadeh, T.  Rostami,    P.V.   Moniz,   Int. J. Mod. Phys. D  \textbf{25}, 1630009 (2016).

\bibitem{shahram} S. Jalalzadeh, A.J.S. Capistrano, P.V. Moniz, Phys. Dark Univ.  \textbf{18},  55 (2017), arXiv:1709.09923.

{\bibitem{Blyth} T. Dereli, M. Panahimoghaddam, R.W. Tucker, Journal of Physics A \textbf{15},  3167 (1982);
  W.F. Blyth, C.J. Isham , Phys. Rev. D \textbf{11}, 768 (1975);
  K. Enqvist, K.W. Ng, K.A. Olive, Nucl. Phys. B \textbf{303}, 713 (1988).}


{\bibitem{Kontsevich}  M. Kontsevich, Lett. Math. Phys. \textbf{48}, 35 (1999), arXiv:math/9904055.}

\bibitem{wigner}  T. Curtright, D. Fairlie, C. Zachos, Phys. Rev. D \textbf{58}, 025002 (1998), arXiv:hep-th/9711183;
  F. Bonneau, M. Gerstenhaber, A. Giaquinto, D. Sternheimer, J. Math. Phys. \textbf{45}, 3703 (2004).
{ \bibitem{Bopp} F. Bopp, Ann. Inst. H. Poincar\'e \textbf{15}, 81 (1956).}
\bibitem{Bojo} M. Bojoward, Phys. Rev. D \textbf{64}, 084018 (2001).
\bibitem{DeWitt1} B.S. DeWitt, in: Relativity, (ed. M. Carmeli and S. I. Fickler), New York (1970).
\bibitem{Tipler} F.J. Tipler, Phys. Rep. \textbf{137}, 231 (1986).

\bibitem{Tucker}  T. Dereli et al, Class. Quantum Grav. \textbf{10}, 1425  (1993).
\bibitem{Shlomo} S. Shlomo and M. Prakash, Nucl. Phys. A \textbf{357}, 157 (1981).

 \bibitem{Cordero} R. Cordero, H. Garc\'ia-Compe\'an, F.J. Turrubiates Phys. Rev. D \textbf{83}, 125030 (2011), arXiv:1302.3117.

\bibitem{Packet} S. Jalalzadeh, F. Ahmadi, H.R. Sepangi, J. High Energy Phys. {\bf08}, 012 (2003); P. Pedram, S. Jalalzadeh, Phys. Lett. B {\bf660}, 1 (2008); P. Pedram,  M. Mirzaei,  S. Jalalzadeh, S.S. Gousheh, Gen. Relativ. Gravit. {\bf40}, 1663 (2008); S. Jalalzadeh, B. Vakili, Int. J. Theor. Phys. {\bf51}, 263 (2012).

\bibitem{Berry} M.V. Berry,  Philos. Trans. R. Soc. London \textbf{287}, 237 (1977).

\bibitem{Tatar} V. I. Tatarskii, Usp. Fiz. Nauk. {\bf139}, 587 (1983) [Sov. Phys. Usp. {\bf26}, 311 (1983)].
\bibitem{Koe} R. Koekoek, J. Math. Anal. Appl. {\bf153}, 576 (1990).


\bibitem{Dirac} P. A. M. Dirac, Lectures on Quantum Mechanics (Belfere
Graduate School of Science, Yeshiva University Press, New York, 1964).
\bibitem{rashki} M. Rashki, S. Jalalzadeh,  Phys. Rev. D \textbf{91}, 023501 (2015), arXiv:1412.3950.

\bibitem{rashki2} M. Rashki, S. Jalalzadeh, Gen. Rel. Grav. \textbf{49}, 14  (2017), arXiv:1612.03047.

\bibitem{Timepass} A. Barvinsky, Phys. Rep. \textbf{230}, 237 (1993; S.A. Gogilidze, A. M. Khvedelidze, V.V. Papoyan, Yu.G. Palii, V.N. Pervushin, Grav. Cosmol. \textbf{3}, 17 (1997); A.M. Khvedelidze, V.V. Papoyan, Yu.G. Palii, V.N. Pervushin, Phys. Lett. B \textbf{402}, 263 (1997);  H. Farajollahi, Int. J. Theor. Phys. \textbf{47}, 1479 (2008).




\bibitem{Robert} R.A. Leaeock, Found. Phys. \textbf{17}, 799 (1987).
\bibitem{Oliveira} C.R de Oliveira, C.P. Malta, Ann. Phys. \textbf{155}, 447 (1984).

\bibitem{Zac} C Zachos, T Curtright, Prog. Theor. Phys. Suppl. \textbf{135}, 244 (1999), arXiv:hep-th/9903254.





















\end{thebibliography}

\end{document}